\documentclass{birkjour}

\usepackage[T2A]{fontenc}
\usepackage[english]{babel}
\usepackage{latexsym}
\usepackage{amssymb}
\usepackage{amsmath}

 \newtheorem{thm}{Theorem}[section]
 \newtheorem{cor}[thm]{Corollary}

 \theoremstyle{definition}
 
 \theoremstyle{remark}
 \newtheorem{rem}[thm]{Remark}
 
 \numberwithin{equation}{section}

\def\lemma #1. #2\par{\medbreak
  \noindent{\tt {\bf Lemma #1.}\enspace}{\sl#2\par}%
  \ifdim\lastskip<\medskipamount \removelastskip\penalty55\medskip\fi}
\def\sledstv #1. #2\par{\medbreak
  \noindent{\tt {\bf Corollary #1.}\enspace}{\sl#2\par}%
  \ifdim\lastskip<\medskipamount \removelastskip\penalty55\medskip\fi}

\def\{{\lbrace}
\def\}{\rbrace}

\def\cl{{\mathbb C}\!\ell}

\def\ss{\stackrel}

\def\L{{\cal L}}

\def\O{{\rm O}}

\def\exp{{\rm exp}}
\def\cos{{\rm cos}}
\def\sin{{\rm sin}}
\def\be{\begin{equation}}
\def\ee{\end{equation}}

\newcommand{\st}{\stackrel}
\newcommand{\vsp}{{\vrule width0pt height15pt}}

\begin{document}

\title[Development of the method of quaternion typification]{Development of the method of quaternion typification of Clifford algebra elements}

\author[D.~S.~Shirokov]{D.~S.~Shirokov}
\address{Steklov Mathematical Institute\\
Gubkin St.8, 119991 Moscow, Russia}
\email{shirokov@mi.ras.ru}
\begin{abstract}
In this paper we further develop the method of quaternion typification of Clifford algebra elements suggested by the author in the previous papers. On the basis of new classification of Clifford algebra elements, it is possible to reveal and prove a number of new properties of Clifford algebras. We use k-fold commutators and anticommutators. In this paper we consider Clifford and exterior degrees and elementary functions of Clifford algebra elements.
\end{abstract}
\maketitle

\section{Introduction}
In this paper we develop the method of quaternion typification of Clifford algebra elements. This method was suggested in \cite{quattyp}. On the basis of new classification of Clifford algebra elements, it is possible to reveal and prove a number of new properties of Clifford algebras. In our work we use k-fold commutators and anticommutators. Also we consider Clifford and exterior degrees and elementary functions of Clifford algebra elements.
We develop results of \cite{quattyp} and use results of \cite{Marchuk:Shirokov} and \cite{Shirokov}.

\section{Main ideas of the method of quaternion typification of Clifford algebra elements}

Let $p, q$ be nonnegative integers such that $p+q=n$, $n\geq1$. We consider the Clifford algebra $\cl(p,q)$ over the fields of real or complex numbers. A detailed description of the construction of the Clifford algebra $\cl(p,q)$ is given in \cite{Lounesto} and \cite{Marchuk:Shirokov}. Let $e$ be the identity element and let $e^a$, $a=1,\ldots,n$ be generators of the Clifford algebra $\cl(p,q)$,
$$
e^a e^b+e^b e^a=2\eta^{ab}e,
$$
where $\eta=||\eta^{ab}||$ is the diagonal matrix, which diagonal contains $p$ elements equal to $+1$ and $q$ elements equal to $-1$. The elements
$$
e^{a_1\ldots a_k}=e^{a_1}\ldots e^{a_k},\quad a_1<\ldots<a_k,\,k=1,\ldots,n
$$
together with the identity element $e$ form the basis in the Clifford
algebra. The number of basis elements equals $2^n$.

We denote the vector spaces spanned by the elements $e^{a_1\ldots a_k}$ by $\cl_k(p,q)$. The elements of $\cl_k(p,q)$ are denoted by $\st{k}{U}$ and called elements of rank $k$.

Also we have classification of Clifford algebra elements based on the notion of pariry:
\begin{eqnarray}
\cl(p,q)=\cl_{even}(p,q)\oplus\cl_{odd}(p,q),\label{evenness}
\end{eqnarray}
where
$$\cl_{even}(p,q)=\cl_0(p,q)\oplus\cl_2(p,q)\oplus\cl_4(p,q)\oplus\ldots$$
$$\cl_{odd}(p,q)=\cl_1(p,q)\oplus\cl_3(p,q)\oplus\cl_5(p,q)\oplus\ldots$$

Denote by $[U,V]$ the commutator and  by $\{U,V\}$ the anticommutator of
Clifford algebra elements $U, V \in \cl(p,q)$
\begin{eqnarray}
[U,V]=UV-VU,\quad \{U,V\}=UV+VU \label{comanticom}
\end{eqnarray}
and note that
\begin{eqnarray}
UV=\frac{1}{2}[U,V]+\frac{1}{2}\{U,V\}.\label{proizv}
\end{eqnarray}

\bigskip

Consider the Clifford algebra as the vector space and represent it in the form of the direct sum of four subspaces:
\begin{equation}
\cl(p,q)=\cl_{\overline 0}(p,q)\oplus\cl_{\overline 1}(p,q)\oplus
\cl_{\overline 2}(p,q)\oplus\cl_{\overline 3}(p,q),\label{kv}
\end{equation}
where
\begin{eqnarray*}
\cl_{\overline
0}(p,q)&=&\cl_0(p,q)\oplus\cl_4(p,q)\oplus\cl_8(p,q)\oplus\ldots,\\
\cl_{\overline
1}(p,q)&=&\cl_1(p,q)\oplus\cl_5(p,q)\oplus\cl_9(p,q)\oplus\ldots,\\
\cl_{\overline
2}(p,q)&=&\cl_2(p,q)\oplus\cl_6(p,q)\oplus\cl_{10}(p,q)\oplus\ldots,\\
\cl_{\overline
3}(p,q)&=&\cl_3(p,q)\oplus\cl_7(p,q)\oplus\cl_{11}(p,q)\oplus\ldots
\end{eqnarray*}
and the right-hand sides are direct sums of subspaces whose dimensions differ by 4. We suppose that $\cl_k(p,q)=\emptyset$ for $k>p+q$.

We use following notations:
$$
\cl_{\overline{kl}}(p,q)=\cl_{\overline
k}(p,q)\oplus\cl_{\overline l}(p,q),\quad 0\leq k<l\leq 3.
$$
$$
\cl_{\overline{klm}}(p,q)=\cl_{\overline
k}(p,q)\oplus\cl_{\overline l}(p,q)\oplus\cl_{\overline
m}(p,q),\quad 0\leq k<l<m\leq 3.
$$

Consider elements of the Clifford algebra $\cl(p,q)$ from different subspaces
\begin{eqnarray}
\cl_{\overline 0}(p,q),\quad \cl_{\overline 1}(p,q),\quad \cl_{\overline 2}(p,q),\quad \cl_{\overline 3}(p,q),\quad \cl_{\overline {01}}(p,q),\quad \cl_{\overline {02}}(p,q),\nonumber \\
\cl_{\overline {03}}(p,q),\quad \cl_{\overline {12}}(p,q), \quad \cl_{\overline {13}}(p,q),\quad \cl_{\overline {23}}(p,q),\quad \cl_{\overline {012}}(p,q),\label{tip}\\
\cl_{\overline {013}}(p,q),\quad \cl_{\overline {023}}(p,q),\quad \cl_{\overline {123}}(p,q),\quad
\cl_{\overline {0123}}(p,q)=\cl(p,q).\nonumber
\end{eqnarray}

We say that these elements have different {\it quaternion types} (or {\it types}). Elements of subspaces $\cl_{\overline 0}(p,q),\, \cl_{\overline 1}(p,q),\, \cl_{\overline 2}(p,q),\, \cl_{\overline 3}(p,q)$ are called {\it elements of the main quaternion types}. Elements of other types are represented in the form of sums of elements of the main quaternion types. Suppose that the zero element of the Clifford algebra $\cl(p,q)$ belongs to any quaternion type. So, we have classification of elements of the Clifford algebra $\cl(p,q)$  on 15 quaternion types.

We denote elements of quaternion type $\overline k$ by $\st{\overline k}{U}$. Sometimes we denote subspace $\cl_{\overline k}(p,q)$ by $\overline{\textbf{k}}$ and any Clifford algebra element $\st{\overline k}{U}\in\cl_{\overline k}(p,q)$ by $\overline{k}$. When we write ``quaternion type $\overline{k}$'' we mean by $\overline{k}$ a symbol of quaternion type (not an Clifford algebra element). Then
$[\overline k,\overline l]\subseteq\overline{\textbf{m}}$ means that commutator of any two Clifford algebra elements of quaternion types $\overline k$ and $\overline l$ belongs to subspace $\overline{\textbf{m}}=\cl_{\overline m}(p,q)$.

We have the following properties (see \cite{quattyp}):
\begin{eqnarray}
&&[\overline k,\overline k]\subseteq\overline{\textbf{2}},\qquad k=0, 1, 2, 3; \nonumber\\
&&[\overline k,\overline 2]\subseteq\overline{\textbf{k}}, \qquad k=0, 1, 2, 3; \label{1} \\
&&[\overline 0,\overline 1]\subseteq\overline{\textbf{3}}, \quad  [\overline 0,\overline 3]\subseteq\overline{\textbf{1}}, \quad [\overline 1,\overline 3]\subseteq\overline{\textbf{0}} \nonumber,
\end{eqnarray}
\begin{eqnarray}
&&\{\overline k,\overline k\}\subseteq\overline{\textbf{0}},\qquad k=0, 1, 2, 3; \nonumber\\
&&\{\overline k,\overline 0\}\subseteq\overline{\textbf{k}}, \qquad k=0, 1, 2, 3; \label{2} \\
&&\{\overline 1,\overline 2\}\subseteq\overline{\textbf{3}},  \quad \{\overline 1,\overline 3\}\subseteq\overline{\textbf{2}}, \quad \{\overline 2,\overline 3\}\subseteq\overline{\textbf{1}}\nonumber.
\end{eqnarray}

\section{Threefold commutators and anticommutators}
Let's define {\it threefold commutator} and {\it threefold anticommutator} of three elements $U, V, W$ of Clifford algebra $\cl(p,q)$:
\begin{eqnarray}
[U,V,W]&=&UVW-WVU,\\
\{U,V,W\}&=&UVW+WVU.\nonumber
\end{eqnarray}
Using (\ref{comanticom}), we get
$$[[U,V],W]=UVW-VUW-WUV+WVU=\{U,V,W\}-\{V,U,W\},$$
$$\{\{U,V\},W\}=UVW+VUW+WUV+WVU=\{U,V,W\}+\{V,U,W\},$$
$$\{[U,V],W\}=UVW-VUW+WUV-WVU=[U,V,W]-[V,U,W],$$
$$[\{U,V\},W]=UVW+VUW-WUV-WVU=[U,V,W]+[V,U,W],$$
\begin{equation}[U,V,W]=\frac{1}{2}(\{[U,V],W\}+[\{U,V\},W]),\label{trcom}\end{equation}
$$\{U,V,W\}=\frac{1}{2}([[U,V],W]+\{\{U,V\},W\}),$$
$$UVW=\frac{1}{4}([[U,V],W]+\{\{U,V\},W\}+[\{U,V\},W]+\{[U,V],W\})=$$
$$=\frac{1}{2}([U,V,W]+\{U,V,W\}).$$

\begin{thm}
\begin{enumerate}
  \item For elements $U, V, W \in \cl(p,q)$ of the given quaternion types $\overline k, \overline l, \overline m$ (\ref{tip}) the following elements
 \begin{eqnarray}
 [[\st{\overline k}{U},\st{\overline l}{V}],\st{\overline m}{W}],\quad \{\{\st{\overline k}{U},\st{\overline l}{V}\},\st{\overline m}{W}\},\quad \{\st{\overline k}{U},\st{\overline l}{V},\st{\overline m}{W}\} \label{*}
 \end{eqnarray}
 have the same quaternion type $\overline r$, i.e. in the other notation we have
 \begin{eqnarray}
 [[\overline k, \overline l], \overline m],\quad \{\{\overline k, \overline l\}, \overline m\},\quad \{\overline k, \overline l, \overline m\} \subseteq \overline{\textbf{r}}.\label{*2}
 \end{eqnarray}
 Besides, quaternion type does not depend on transposition of $\st{\overline k}{U}, \st{\overline l}{V}, \st{\overline m}{W}$ in (\ref{*})(or $\overline k, \overline l, \overline m$ in (\ref{*2})).
  \item For elements $U, V, W \in \cl(p,q)$ of the given quaternion types $\overline k, \overline l, \overline m$ (\ref{tip}) the following elements
 \begin{eqnarray}\{[\st{\overline k}{U},\st{\overline l}{V}],\st{\overline m}{W}\},\quad [\{\st{\overline k}{U},\st{\overline l}{V}\},\st{\overline m}{W}],\quad [\st{\overline k}{U},\st{\overline l}{V},\st{\overline m}{W}] \label{**}
 \end{eqnarray}
 have the same quaternion type $\overline r$, i.e.
 \begin{eqnarray}
 \{[\overline k, \overline l], \overline m\},\quad [\{\overline k, \overline l\}, \overline m],\quad [\overline k, \overline l, \overline m]\subseteq \overline{\textbf{r}}.\label{**2}
 \end{eqnarray}
  Besides, quaternion type does not depend on transposition of $\st{\overline k}{U}, \st{\overline l}{V}, \st{\overline m}{W}$ in (\ref{**})(or $\overline k, \overline l, \overline m$ in (\ref{**2})).
\end{enumerate}
\end{thm}

\begin{proof} Using (\ref{1}), (\ref{2}) and considering all variants we get that the following expressions $$[[\st{\overline k}{U},\st{\overline l}{V}],\st{\overline m}{W}]\quad \mbox{and} \quad\{\{\st{\overline k}{U},\st{\overline l}{V}\},\st{\overline m}{W}\},\quad \{[\st{\overline k}{U},\st{\overline l}{V}],\st{\overline m}{W}\}\quad \mbox{and}\quad [\{\st{\overline k}{U},\st{\overline l}{V}\},\st{\overline m}{W}]$$ have the same quaternion type. Further we use (\ref{trcom}).
\end{proof}

\begin{rem} Taking into account the theorem statements, we often speak about quaternion types of only two elements $$\{\st{\overline k}{U},\st{\overline l}{V},\st{\overline m}{W}\}, \quad [\st{\overline k}{U},\st{\overline l}{V},\st{\overline m}{W}],$$ meaning also other four elements considered in the theorem. Quaternion type does not depend on transposition $\st{\overline k}{U}, \st{\overline l}{V}, \st{\overline m}{W}$ in the considered threefold commutators and anticommutators.
\end{rem}

Let's write down quaternion types of considered expressions in the case of the main quaternion types:
\bigskip

$$\begin{tabular}{|c|c|c|c|c|c||c|c|c|c|c|c|}
\hline\vsp
$\overline k$ & $\overline l$ & $\overline m$ & $\{\overline k,\overline l,\overline m\}$ & $[\overline k,\overline l,\overline m]$ & $\overline k\overline l\overline m$ &$\overline k$ & $\overline l$ & $\overline m$ & $\{\overline k,\overline l,\overline m\}$ & $[\overline k,\overline l,\overline m]$ & $\overline k\overline l\overline m$ \\ \hline\hline\vsp
$\overline 0$ & $\overline 0$ & $\overline 0$ & $\overline 0$ & $\overline 2$ & $\overline {02}$  &$\overline 1$ & $\overline 1$ & $\overline 1$ & $\overline 1$ & $\overline 3$ & $\overline {13}$\\
$\overline 0$ & $\overline 0$ & $\overline 1$ & $\overline 1$ & $\overline 3$ & $\overline {13}$  &$\overline 1$ & $\overline 1$ & $\overline 2$ & $\overline 2$ & $\overline 0$ & $\overline {02}$\\
$\overline 0$ & $\overline 0$ & $\overline 2$ & $\overline 2$ & $\overline 0$ & $\overline {02}$  &$\overline 1$ & $\overline 1$ & $\overline 3$ & $\overline 3$ & $\overline 1$ & $\overline {13}$\\
$\overline 0$ & $\overline 0$ & $\overline 3$ & $\overline 3$ & $\overline 1$ & $\overline {13}$  &$\overline 1$ & $\overline 2$ & $\overline 2$ & $\overline 1$ & $\overline 3$ & $\overline {13}$\\
$\overline 0$ & $\overline 1$ & $\overline 1$ & $\overline 0$ & $\overline 2$ & $\overline {02}$  &$\overline 1$ & $\overline 2$ & $\overline 3$ & $\overline 0$ & $\overline 2$ & $\overline {02}$\\
$\overline 0$ & $\overline 1$ & $\overline 2$ & $\overline 3$ & $\overline 1$ & $\overline {13}$  &$\overline 1$ & $\overline 3$ & $\overline 3$ & $\overline 1$ & $\overline 3$ & $\overline {13}$\\
$\overline 0$ & $\overline 1$ & $\overline 3$ & $\overline 2$ & $\overline 0$ & $\overline {02}$  &$\overline 2$ & $\overline 2$ & $\overline 2$ & $\overline 2$ & $\overline 0$ & $\overline {02}$\\
$\overline 0$ & $\overline 2$ & $\overline 2$ & $\overline 0$ & $\overline 2$ & $\overline {02}$  &$\overline 2$ & $\overline 2$ & $\overline 3$ & $\overline 3$ & $\overline 1$ & $\overline {13}$\\
$\overline 0$ & $\overline 2$ & $\overline 3$ & $\overline 1$ & $\overline 3$ & $\overline {13}$  &$\overline 2$ & $\overline 3$ & $\overline 3$ & $\overline 2$ & $\overline 0$ & $\overline {02}$\\
$\overline 0$ & $\overline 3$ & $\overline 3$ & $\overline 0$ & $\overline 2$ & $\overline {02}$  &$\overline 3$ & $\overline 3$ & $\overline 3$ & $\overline 3$ & $\overline 1$ & $\overline {13}$\\ \hline
\end{tabular}$$

\section{Some formulas for quaternion types of 2-fold and 3-fold commutators and anticommutators in the cases of main quaternion types}

From (\ref{1}) and (\ref{2}) we have:
$$
\{\overline k,\overline l\}=\left\lbrace
\begin{array}{ll}
\overline{ (k + l + 2){\rm mod}\,4}, & \parbox{.70\linewidth}{\rm if $k$ and $l$ are odd ;}\\
\overline{ (k + l){\rm mod}\,4}, & \parbox{.70\linewidth}{\rm other cases .}
\end{array}
\right.
$$
$$
[\overline k,\overline l]=\left\lbrace
\begin{array}{ll}
\overline{ (k + l){\rm mod}\,4}, & \parbox{.70\linewidth}{\rm if $k$ and $l$ are odd;}\\
\overline{ (k + l + 2){\rm mod}\,4}, & \parbox{.70\linewidth}{\rm other cases.}
\end{array}
\right.
$$
or
\begin{eqnarray}
[\overline k,\overline l]=\overline{ (k+l+1+(-1)^{kl}){\rm mod}\,4},\label{form3}\\
\{\overline k,\overline l\}=\overline{ (k+l+1-(-1)^{kl}){\rm mod}\,4}.\nonumber
\end{eqnarray}
$$
\st{\overline k}{U}\st{\overline l}{V}\subseteq\left\lbrace
\begin{array}{ll}
\overline{\textbf{02}}, & \mbox{\rm if $k+l$ - even;}\\
\overline{\textbf{13}}, & \mbox{\rm if $k+l$  - odd.}
\end{array}
\right.
$$

Statements of Theorem 3.1 can be written down more compactly in the form of the following formulas, where $\overline k, \overline l, \overline m$ - main quaternion types:
\begin{eqnarray}
[\overline k,\overline l,\overline m]=\overline{ (k+l+m+1+(-1)^{kl+km+lm}){\rm mod}\,4},\label{form2}\\
\{\overline k,\overline l,\overline m\}=\overline{ (k+l+m+1-(-1)^{kl+km+lm}){\rm mod}\,4}.\nonumber
\end{eqnarray}
$$
\st{\overline k}{U}\st{\overline l}{V}\st{\overline m}{W}\subseteq\left\lbrace
\begin{array}{ll}
\overline{\textbf{02}}, & \mbox{\rm if $k+l+m$ - even;}\\
\overline{\textbf{13}}, & \mbox{\rm if $k+l+m$  - odd.}
\end{array}
\right.
$$

Below we give generalization of these formulas on $k$-fold commutators and $k$-fold anticommutators.

Further it will be not always convenient to use formulas (\ref{form2}) and (\ref{form3}). Therefore, let us consider operations {\em sharp} $\sharp$, {\em flat} $\flat$ and {\em natural} $\natural$ with the following properties:
\begin{eqnarray}
&(\overline 0)_\sharp&=\overline 2,\quad  (\overline 1)_\sharp=\overline 3,\quad (\overline 2)_\sharp=\overline 0,\quad (\overline 3)_\sharp=\overline 1,\nonumber\\
&(\overline 0)_\flat&=\overline 1,\quad (\overline 1)_\flat=\overline 0,\quad (\overline 2)_\flat=\overline 3,\quad (\overline 3)_\flat=\overline 2,\\
&(\overline 0)_\natural&=\overline 3,\quad (\overline 1)_\natural=\overline 2,\quad (\overline 2)_\natural=\overline 1,\quad (\overline 3)_\natural=\overline 0,\nonumber
\end{eqnarray}
Note that
\begin{eqnarray}
&&({\overline k}_\sharp)_\sharp=\overline k,\quad ({\overline k}_\flat)_\flat=\overline k,\quad ({\overline k}_\natural)_\natural=\overline k,\nonumber\\
&&\overline k_\sharp=\overline{(k +2){\rm mod}\,4},\quad \overline k_\flat=\overline{(k_\natural +2){\rm mod}\,4},\nonumber\\
&&\overline k_\flat=\overline{(k+(-1)^k){\rm mod}\,4}=\overline{(k+2-(-1)^k){\rm mod}\,4},\nonumber\\
&&\overline k_\natural=\overline{(k-(-1)^k){\rm mod}\,4}=\overline{(k+2+(-1)^k){\rm mod}\,4}.\nonumber
\end{eqnarray}
By Theorem 3.1 and by these formulas, it follows that
$$\exists \overline{\textbf{r}} : [[\overline k, \overline l], \overline m]_\sharp, \{\{\overline k, \overline l\}, \overline m\}_\sharp, \{\overline k, \overline l, \overline m\}_\sharp , \{[\overline k, \overline l], \overline m\}, [\{\overline k, \overline l\}, \overline m], [\overline k, \overline l, \overline m]\subseteq\overline{\textbf{r}}.$$

Let us consider trivial (identity) operation $I$ with the property $(\overline k)_I=\overline k$. Then the
following properties for operations $I, \sharp, \flat, \natural$ are fulfilled:
\begin{eqnarray}
I \circ I = \sharp \circ \sharp = \flat \circ \flat = \natural \circ \natural = I, \nonumber\\
I \circ \sharp = \sharp \circ I = \flat \circ \natural = \natural \circ \flat = \sharp, \\
I \circ \flat = \flat \circ I = \sharp \circ \natural = \natural \circ \sharp = \flat, \nonumber \\
I \circ \natural = \natural \circ I = \sharp \circ \flat = \flat \circ \sharp = \natural. \nonumber
\end{eqnarray}
These properties are similar to the properties of quaternions with respect to the operation of composition $\circ$.

Consider 3-fold commutators and anticommutators with 2 given main quaternion types. Then we have:
\begin{eqnarray}
&\{\overline 0, \overline 0, \overline k \},&\{\overline 1, \overline 1, \overline k \},\ \{\overline 2, \overline 2,\ \overline k \},\ \{\overline 3, \overline 3, \overline k \}\subseteq\overline{\textbf{k}},\nonumber\\
&\{\overline 0, \overline 2, \overline k \},&\{\overline 1, \overline 3, \overline k \}\subseteq\overline{\textbf{k}}_\sharp,\\\label{simv1}
&\{\overline 0, \overline 1, \overline k \},&\{\overline 2, \overline 3, \overline k \}\subseteq\overline{\textbf{k}}_\flat\nonumber,\\
&\{\overline 0, \overline 3, \overline k \},&\{\overline 1, \overline 2, \overline k \}\subseteq\overline{\textbf{k}}_\natural,\nonumber
\end{eqnarray}
\begin{eqnarray}
&[\overline 0, \overline 2, \overline k ],& [\overline 1, \overline 3, \overline k ]\subseteq \overline{\textbf{k}}\nonumber,\\
&[\overline 0, \overline 0, \overline k ],& [\overline 1, \overline 1, \overline k ],\ [\overline 2, \overline 2, \overline k ],\ [\overline 3, \overline 3, \overline k ]\subseteq\overline{\textbf{k}}_\sharp,\\\label{simv2}
&[\overline 0, \overline 3, \overline k ],& [\overline 1, \overline 2, \overline k ]\subseteq\overline{\textbf{k}}_\flat\nonumber,\\
&[\overline 0, \overline 1, \overline k ],& [\overline 2, \overline 3, \overline k ]\subseteq\overline{\textbf{k}}_\natural.\nonumber
\end{eqnarray}
For 2-fold commutators and anticommutators we have the following formulas:
\begin{eqnarray}
&\{\overline k, \overline k \}&\subseteq\overline{\textbf{0}}, \quad
\{\overline k, \overline 0 \}\subseteq\overline{\textbf{k}},\\\label{2simv1}
&\{\overline k, \overline 2 \}&\subseteq\overline{\textbf{k}}_\sharp,\quad
\{\overline k, \overline 1 \}\subseteq\overline{\textbf{k}}_\flat,\quad
\{\overline k, \overline 3 \}\subseteq\overline{\textbf{k}}_\natural,\nonumber
\end{eqnarray}
\begin{eqnarray}
&[\overline k, \overline k ]&\subseteq\overline{\textbf{2}},\quad
[\overline k, \overline 2 ]\subseteq\overline{\textbf{k}},\\\label{2simv2}
&[\overline k, \overline 0 ]&\subseteq\overline{\textbf{k}}_\sharp,\quad
[\overline k, \overline 3 ]\subseteq\overline{\textbf{k}}_\flat,\quad
[\overline k, \overline 1 ]\subseteq\overline{\textbf{k}}_\natural.\nonumber
\end{eqnarray}
So, for the main quaternion types we have:
$$\{\overline l, \overline l, \overline k \},\ [\overline 0, \overline 2, \overline k ],\ [\overline 1, \overline 3, \overline k ]\subseteq\overline{\textbf{k}},$$
$$[\overline l, \overline l, \overline k ],\ \{\overline 0, \overline 2, \overline k \},\ \{\overline 1, \overline 3, \overline k \}\subseteq\overline{\textbf{k}}_\sharp.$$

\section{$k$-fold commutators and anticommutators}

Let us give definitions for {\it $k$-fold commutator} and {\it $k$-fold anticommutator} of $k$ elements $U_1, U_2, \ldots, U_k$ of Clifford algebra $\cl(p,q)$:
\begin{eqnarray}
[U_1, U_2, \ldots, U_k]=U_1 U_2 \ldots U_k-U_k \ldots U_2 U_1;\\
\{U_1, U_2, \ldots, U_k\}=U_1 U_2 \ldots U_k+U_k \ldots U_2 U_1.
\end{eqnarray}
Then
\begin{eqnarray}
U_1 U_2 \ldots U_k=\frac{1}{2}([U_1, U_2, \ldots, U_k]+\{U_1, U_2, \ldots, U_k\}).\label{kcom0}
\end{eqnarray}
\begin{thm} For elements $U_1, U_2, \ldots, U_k \in \cl(p,q)$ the following formulas are fulfilled:
\begin{eqnarray}
&U_1 U_2 \ldots U_k&=\frac{1}{2^{k-1}}\sum_{j=1}^{2^{k-1}}d_j^{k-1},\label{kcom1}\\
&[U_1, U_2, \ldots, U_k]&=\frac{1}{2^{k-2}}\sum_{j=1}^{2^{k-2}}d_j^{{k-1}^+},\label{kcom2}\\
&\{U_1, U_2, \ldots, U_k\}&=\frac{1}{2^{k-2}}\sum_{j=1}^{2^{k-2}}d_j^{{k-1}^-},\label{kcom3}
\end{eqnarray}
where $d_j^{k-1}$ ($j$ from $1$ to $2^{k-1}$) - all possible expressions of the form $$\underbrace{(\ldots((}_{k-1}U_1,U_2),U_3),\ldots,U_k),$$
where bracket $"("$ is $"["$ or $"\{"$, $d_j^{{k-1}^+}$ - expressions $d_j^{k-1}$, where we have odd number of commutators, $d_j^{{k-1}^-}$ - expressions $d_j^{k-1}$, where we have even number of commutators.\\
For Clifford algebra elements $U_1, U_2, \ldots, U_k$ of given main quaternion types $\overline{ a_1}, \overline{a_2}, \ldots, \overline{a_k}$ all expressions $d_j^{{k-1}^+}$ have the same quaternion type. This type is a type (by (\ref{kcom2})) of expression $[U_1, U_2, \ldots, U_k]$ and equals
\begin{equation}
[\overline{ a_1}, \overline{a_2}, \ldots, \overline{a_k}]=\overline{(a_1+a_2+\ldots+a_k+1+(-1)^{\sum_{i<j}^{k}a_i a_j}){\rm mod}\,4}.\label{kcom4}
\end{equation}
For Clifford algebra elements $U_1, U_2, \ldots U_k$ of given main quaternion types $\overline{ a_1}, \overline{a_2}, \ldots, \overline{a_k}$ all expressions $d_j^{{k-1}^-}$ have the same quaternion type. This type is a type (by (\ref{kcom3})) of expression $\{U_1, U_2, \ldots, U_k\}$ and equals
\begin{equation}
\{\overline{ a_1}, \overline{a_2}, \ldots, \overline{a_k}\}=\overline{(a_1+a_2+\ldots+a_k+1-(-1)^{\sum_{i<j}^{k}a_i a_j}){\rm mod}\,4}.\label{kcom5}
\end{equation}
For Clifford algebra elements $U_1, U_2, \ldots, U_k$ of given main quaternion types $\overline{ a_1}, \overline{a_2}, \ldots, \overline{a_k}$ the following formula is fulfilled:
\begin{equation}
\st{\overline{ a_1}}{U_1} \st{\overline{a_2}}{U_2} \ldots \st{\overline{a_k}}{U_k} \subseteq\left\lbrace
\begin{array}{ll}
\overline{\textbf{02}}, & \mbox{\rm if $\sum_{j=1}^k a_j$ - even;}\label{kcom6}\\
\overline{\textbf{13}}, & \mbox{\rm if $\sum_{j=1}^k a_j$  - odd.}
\end{array}
\right.
\end{equation}
\end{thm}

\begin{proof} We use the method of mathematical induction. Formula (\ref{trcom}) is a partial case of formulas (\ref{kcom1}), (\ref{kcom2}), (\ref{kcom3}) in case $k=3$. Suppose that formulas (\ref{kcom1}), (\ref{kcom2}), (\ref{kcom3}) are valid for $k>3$. Let us prove the validity of these formulas for $k+1$:
$$U_1 U_2 \ldots U_k U_{k+1}=\frac{1}{2^{k-1}}(\sum_{j=1}^{2^{k-1}}d_j^{k-1})U_{k+1}=$$
$$=\frac{1}{2^{k-1}}(\frac{1}{2}[\sum_{j=1}^{2^{k-1}}d_j^{k-1},U_{k+1}]+\frac{1}{2}\{\sum_{j=1}^{2^{k-1}}d_j^{k-1},U_{k+1}\})=\frac{1}{2^{k}}(\sum_{j=1}^{2^{k}}d_j^k).$$
The validity of formulas (\ref{kcom2}), (\ref{kcom3}) for $k+1$ follows from
\begin{eqnarray}
&[U_1,U_2,\ldots,U_{k+1}]&=\frac{1}{2}\{[U_1,U_2,\ldots, U_k],U_{k+1}\}+\frac{1}{2}[\{U_1,U_2,\ldots, U_k\},U_{k+1}],\nonumber\\
&\{U_1,U_2,\ldots,U_{k+1}\}&=\frac{1}{2}[[U_1,U_2,\ldots, U_k],U_{k+1}]+\frac{1}{2}\{\{U_1,U_2,\ldots, U_k\},U_{k+1}\}.\nonumber
\end{eqnarray}

Let us prove that all expressions $d_j^{{k-1}^+}$ have quaternion type
$$\overline{(a_1+a_2+\ldots+a_k+1+(-1)^{\sum_{i<j}^{k}a_i a_j}){\rm mod}\,4}.$$
This proves (\ref{kcom4}). Suppose that it is valid for $d_j^{{k-1}^+}$. Consider $d_j^{{k}^+}$. This expression will be $d_j^{{k}^+}=\{d_j^{{k-1}^+},U_{k+1}\}$ or $d_j^{{k}^+}=[d_j^{{k-1}^-},U_{k+1}]$. Let us calculate quaternion types of these 2 expressions using corresponding expressions for $k=2$. We can get the following expressions:

\begin{eqnarray}
&&\{\overline{(\sum_{l=1}^{k}a_l+1+(-1)^{\sum_{i<j}^{k}a_i a_j}){\rm mod}\,4}, \overline{a_{k+1}}\}=\nonumber\\
&&=\overline{(\sum_{l=1}^{k+1}a_l+1+(-1)^{\sum_{i<j}^{k}a_i a_j}+1-(-1)^{a_{k+1}\sum_{l=1}^{k}a_l+2a_{k+1}}){\rm mod}\,4}=\nonumber\\
&&=\overline{(\sum_{l=1}^{k+1}a_l+1+(-1)^{\sum_{i<j}^{k+1}a_i a_j})){\rm mod}\,4};\nonumber
\end{eqnarray}
\begin{eqnarray}
&&[\overline{(\sum_{l=1}^{k}a_l+1-(-1)^{\sum_{i<j}^{k}a_i a_j}){\rm mod}\,4}, \overline{a_{k+1}}]=\nonumber\\
&&=\overline{(\sum_{l=1}^{k+1}a_l+1-(-1)^{\sum_{i<j}^{k}a_i a_j}+1+(-1)^{a_{k+1}\sum_{l=1}^{k}a_l+2a_{k+1}}){\rm mod}\,4}=\nonumber\\
&&=\overline{(\sum_{l=1}^{k+1}a_l+1+(-1)^{\sum_{i<j}^{k+1}a_i a_j})){\rm mod}\,4}.\nonumber
\end{eqnarray}

The proof of (\ref{kcom5}) is analogous. Using (\ref{kcom4}), (\ref{kcom5}) and (\ref{kcom0}), we get (\ref{kcom6}).
\end{proof}

Let's write down the statement of Theorem 5.1 for $k=4$ ($A, B, C, D$ - elements of Clifford algebra $\cl(p,q$)):
\begin{eqnarray}
&ABCD&=\frac{1}{8}([[[A,B],C],D]+[\{\{A,B\},C\},D]+[[\{A,B\},C],D]+\nonumber\\
&&+[\{[A,B],C\},D]+\{[[A,B],C],D\}+\{\{\{A,B\},C\},D\}+\nonumber\\
&&+\{[\{A,B,\},C],D\}+\{\{[A,B],C\},D\}).\nonumber
\end{eqnarray}
\begin{eqnarray}
&[A,B,C,D]&=\frac{1}{4}([[[A,B],C],D]+[\{\{A,B\},C\},D]+\nonumber\\
&&+\{[\{A,B,\},C],D\}+\{\{[A,B],C\},D\}).\nonumber
\end{eqnarray}
\begin{eqnarray}
&\{A,B,C,D\}&=\frac{1}{4}([[\{A,B\},C],D]+[\{[A,B],C\},D]+\nonumber\\
&&+\{[[A,B],C],D\}+\{\{\{A,B\},C\},D\}.\nonumber
\end{eqnarray}
\begin{equation}
[\overline{ a}, \overline{b}, \overline{c}, \overline{d}]=\overline{(a+b+c+d+1+(-1)^{ab+ac+ad+bc+bd+cd}){\rm mod}\,4}.\nonumber
\end{equation}
\begin{equation}
\{\overline{ a}, \overline{b}, \overline{c}, \overline{d}\}=\overline{(a+b+c+d+1-(-1)^{ab+ac+ad+bc+bd+cd}){\rm mod}\,4}.\nonumber
\end{equation}
\begin{equation}
\st{\overline{a}}{A} \st{\overline{b}}{B}\st{\overline{c}}{C}\st{\overline{d}}{D} =\left\lbrace
\begin{array}{ll}
\st{\overline{02}}{W}, & \mbox{\rm if $a+b+c+d$ - even;}\nonumber\\
\st{\overline{13}}{W}, & \mbox{\rm if $a+b+c+d$  - odd.}
\end{array}
\right.
\end{equation}

\begin{rem} Formula (\ref{kcom6}) displays quaternion type of product of any number of Clifford algebra elements. Note that this formula doesn't present anything new since it follows from lemma \cite{Hestenes}:
$$\st{k}{U}\st{l}{V}=\st{k-l}{W}+\st{k-l+2}{W}+\ldots+\st{k+l}{W}, \quad \mbox{where}\st{m}{W}=0\quad \mbox{for} \quad m>n\quad \mbox{or}\quad m<0.$$
\end{rem}

Consider two Clifford algebra elements $U, V$ of the given quaternion types $\overline k, \overline l$ respectively. Statement $\st{\overline l}{V}\st{\overline k}{U}\st{\overline l}{V}\st{\overline k}{U}\subseteq\overline{\textbf{02}}$ is clear without method of quaternion typification. But for similar expression we have
$$\st{\overline k}{U}\st{\overline l}{V}\st{\overline l}{V}\st{\overline k}{U}\subseteq\overline{\textbf{0}}.$$
It so, because $UVVU=\frac{1}{2}([U,V,V,U]+\{U,V,V,U\})=\frac{1}{2}\{U,V,V,U\}$ and quaternion type of this expression is $\overline{\textbf{0}}$ (see (\ref{kcom5})).

So, now we are able to calculate quaternion type of product $U_1 U_2 \ldots U_k$ of any number of Clifford algebra elements. We can use formula (\ref{kcom6}). But in some cases it's more sensible to calculate at first quaternion type of $k$-fold commutator $[U_1, U_2, \ldots, U_k]$ and $k$-fold anticommutator $\{U_1, U_2, \ldots, U_k\}$, and then take into account (\ref{kcom0}).

\section{Formulas for Clifford and exterior degrees of Clifford algebra elements}

We consider product of Clifford algebra elements in previous sections. This product is called {\it Clifford product}.

Let us define another operation of product:
\begin{equation}
e^{a_1}\wedge e^{a_2}\wedge \ldots \wedge e^{a_k}=e^{[a_1}e^{a_2}\ldots e^{a_k]},
\end{equation}
where square brackets mean operation of alternation of indexes. This product is called {\it exterior product} of Clifford algebra elements. From this definition we get
\begin{eqnarray}
e^{a_1}\wedge e^{a_2}=-e^{a_2}\wedge e^{a_1} \qquad \mbox{for}\quad a_1, a_2 = 1, 2,\ldots,n;\\
e^{a_1}\wedge\ldots \wedge e^{a_k}=e^{a_1}\ldots e^{a_k}=e^{a_1\ldots a_k} \qquad \mbox{for}\quad a_1<\ldots <a_k.
\end{eqnarray}
So, we have Clifford algebra $\cl(p,q)$ with two operation of products (the Clifford product and the exterior product).

We say that {\it Clifford degree} of Clifford algebra element $U\in \cl(p,q)$ is
\begin{equation}
(U)^m=\underbrace{U\vee U\vee U\vee \ldots \vee U}_{m},\label{step}
\end{equation}
where $\vee$ is the Clifford product. We say that {\it exterior degree} of Clifford algebra element $U\in \cl(p,q)$ is
\begin{equation}
(\wedge U)^m=\underbrace{U\wedge U\wedge U\wedge \ldots \wedge U}_{m},\label{vnstep}
\end{equation}
where $\wedge$ is the exterior product.

We are interested in rank or quaternion type of Clifford and exterior degrees of Clifford algebra element.

\begin{thm}
Let $\ss{k}{U}$ be a Clifford algebra $\cl(p,q)$ element of rank $k$:
$$\ss{k}{U}=\sum_{a_1< \ldots <a_k} u_{a_1\ldots a_k} e^{a_1 \ldots a_k}.$$
Then for $m\geq 2$ we have
$$(\wedge \ss{0}{U})^m=u^m e.$$
If $mk\leq n$ and $k$ - even, then 
$$(\wedge \ss{k}{U})^m=\sum(m!)\underbrace{u_{a_1 \ldots a_k} \ldots u_{b_1 \ldots b_k}}_{m} \underbrace{e^{a_1 \ldots a_k}\wedge \ldots \wedge e^{b_1 \ldots b_k}}_{m},$$
where sum is under condition $a_1<\ldots<a_k, \ldots, b_1<\ldots< b_k $ - $(mk)$ different indexes and basis elements $e^{a_1 \ldots a_k}, \ldots, e^{b_1 \ldots b_k}$ are ordered.\\
If $k$ - odd or $mk>n$, then
$$(\wedge \ss{k}{U})^m=0.$$

Thus,
\begin{equation}
(\wedge \ss{k}{U})^m=\left\lbrace
\begin{array}{ll}
\ss{mk}{W},\quad & \mbox{\rm $mk\leq n$ and $k$ - even;}\\ \\
0,\quad & \mbox{\rm other cases.}
\end{array}
\right.\label{th7}
\end{equation}
\end{thm}

\begin{proof} Formula for $(\wedge \ss{0}{U})^m$ is obvious. Further,
\begin{eqnarray*}
&&(\wedge \ss{k}{U})^2=(u_{1\ldots k} e^{1 \ldots k}+ \ldots)\wedge(u_{1\ldots k} e^{1 \ldots k}+ \ldots)=\\
&&=\sum u_{a_1 \ldots a_k} u_{b_1 \ldots b_k} (e^{a_1 \ldots a_k}\wedge e^{b_1 \ldots b_k}+e^{b_1 \ldots b_k}\wedge e^{a_1 \ldots a_k})=\\
&&=\sum u_{a_1 \ldots a_k} u_{b_1 \ldots b_k} (1+(-1)^{k^2})e^{a_1 \ldots a_k}\wedge e^{b_1 \ldots b_k},
\end{eqnarray*}
where $a_1<\ldots < a_k, b_1< \ldots < b_k$ and all basis elements are ordered.
We see that for odd $k$ this expression equals to zero. Hence any exterior degree of Clifford algebra element of odd rank equals to zero.  We get formulas for even $k$ and $m\geq 3$ in the recurrent way from the formula for $m=2$, where $2!=1+(-1)^{k^2}$.
\end{proof}

\begin{thm}
Let $\ss{k}{U}$ be a Clifford algebra $\cl(p,q)$ element of rank $k$. Then for $m\geq 2$ we have
\begin{equation}
(\ss{k}{U})^m=\left\lbrace
\begin{array}{ll}
\ss{0}{W}+\ss{4}{W}+\ldots \ss{km}{W}, & \mbox{\rm $k=4, 8, 12 \ldots$,\quad $m$ - odd ;}\\
\ss{2}{W}+\ss{6}{W}+\ldots \ss{km}{W}, & \mbox{\rm $k=2, 6, 10 \ldots$,\quad $m$ - odd ;}\\
\ss{1}{W}+\ss{5}{W}+\ldots \ss{km-(m-1)}{W}, & \mbox{\rm $k=1, 5, 9 \ldots$,\quad $m$ - odd;}\\
\ss{3}{W}+\ss{7}{W}+\ldots \ss{km-(m-1)}{W}, & \mbox{\rm $k=3, 7, 11 \ldots$,\quad $m$ - odd ;}\\
\ss{0}{W}+\ss{4}{W}+\ldots \ss{km}{W}, & \mbox{\rm $k$ - even,\quad $m$ - even ;}\\
\ss{0}{W}+\ss{4}{W}+\ldots \ss{(k-1)m}{W}, & \mbox{\rm $k$ - odd,\quad $m$ - even .}
\end{array}
\right.\label{th8}
\end{equation}
In the special cases we have
$$(\ss{0}{U})^m=u^m e \in \cl_0(p,q)$$
$$
(\ss{1}{U})^m=\left\lbrace
\begin{array}{ll}
(\sum_{j=1}^{k} (u_j)^2 \eta^{jj})^{\frac{m}{2}} e \in \cl_0(p,q)), & \mbox{\rm $m$ - even ;}\\\\
(\sum_{j=1}^{k} (u_j)^2 \eta^{jj})^{\frac{m-1}{2}} \sum_{i=1}^{k} u_i e^i \in \cl_1(p,q), & \mbox{\rm $m$ - odd.}
\end{array}
\right.
$$
\end{thm}

Let use terminology connected with the method of quaternion typification. Then we can note that

\begin{cor} For Clifford algebra element $U\in \cl(p,q)$ of the given quaternion type $\overline k=\overline 0, \overline 1, \overline 2, \overline 3$ we have
\begin{equation}
(\ss{\overline k}{U})^m=\left\lbrace
\begin{array}{ll}
\ss{\overline k}{W}, & \mbox{\rm $m$ - odd ,}\\
\ss{\overline 0}{W}, & \mbox{\rm $m$ - even.}
\end{array}
\right.\label{zam}
\end{equation}
\end{cor}

\begin{proof} Formula for $(\ss{0}{W})^m$ is obvious. Further,
\begin{eqnarray*}
&&(\ss{1}{U})^2=(u_1 e^1 + u_2 e^2 + \ldots + u_n e^n)(u_1 e^1 + u_2 e^2 + \ldots + u_n e^n)=\\
&&=\sum_{i=1}^{n} (u_i)^2 \eta^{ii} + \sum_{i \neq j} u_i u_j (e^i e^j + e^j e^i)=\sum_{i=1}^{n} (u_i)^2 \eta^{ii}.
\end{eqnarray*}
From this equality we get the formula for $(\ss{1}{U})^m$.
Now let us prove the other statements.
It is clear that $$UU=\frac{1}{2}(UU+UU)=\frac{1}{2}\{U,U\}.$$ Further, $$(U)^4=\frac{1}{8}\{\{U,U\},\{U,U\}\}.$$ If we have $(U)^m$ in the form of anticommutators of Clifford algebra elements for even $m$, then $$(U)^{m+2}=c\{U^m,\{U,U\}\},$$ where $c$ is constant. Then, using $$\{\st{\overline k}{U},\st{\overline k}{U}\}=\st{\overline 0}{U},\qquad k=0, 1, 2, 3,$$ we get the statement of the theorem for even $m$ (see (\ref{2})).
As above, $$(U)^3=\frac{1}{4}\{\{U,U\},U\}.$$ We know the expressions in the form of anticommutators of Clifford algebra elements for even $m$. Then we have $$(U)^{m+1}=c\{U^m,U\},$$ where $c$ is constant.  Using $$\{\st{\overline k}{U},\st{\overline 0}{U}\}=\st{\overline k}{U}, \qquad k=1, 2, 3,$$ we have the statements for odd $m$ (see (\ref{2})).
\end{proof}

\begin{rem} Let take into account the dimension $n$ of Clifford algebra. Then we can make these formulas more precise. For example, for $m=2$ we have 4 different cases (in Theorem 6.2 there are 2 cases). These extra cases appears when dimension of Clifford algebra $n$ is smaller than $km$ (if $mk > n$).
$$
(\ss{k}{U})^2=\left\lbrace
\begin{array}{ll}
\ss{0}{W}+\ss{4}{W}+\ldots \ss{2k}{W}, & \mbox{\rm $n\geq 2k$, $k$ - even ;}\\
\ss{0}{W}+\ss{4}{W}+\ldots \ss{2k-2}{W}, & \mbox{\rm $n\geq 2k$, $k$ - odd ;}\\
\ss{0}{W}+\ss{4}{W}+\ldots \ss{2n-2k}{W}, & \mbox{\rm $n\leq 2k$, $n, k$ are of the same parity ;}\\
\ss{0}{W}+\ss{4}{W}+\ldots \ss{2n-2k-2}{W}, & \mbox{\rm $n\leq 2k$, $n, k$ are of different parity .}
\end{array}
\right.
$$
For $m=8$ we have 12 cases (not 2):
$$
(\ss{k}{U})^4=\left\lbrace
\begin{array}{ll}
\ss{0}{W}+\ss{4}{W}+\ldots \ss{4k}{W}, & \mbox{\rm $k\leq \frac{n}{4}$, $k$ - even ;}\\
\ss{0}{W}+\ss{4}{W}+\ldots \ss{2n-4k}{W}, & \mbox{\rm $\frac{n}{4}\leq k \leq \frac{n}{2}$, $k, n$ - even ;}\\
\ss{0}{W}+\ss{4}{W}+\ldots \ss{2n-4k-2}{W}, & \mbox{\rm $\frac{n}{4}\leq k \leq \frac{n}{2}$, $k$ - even, $n$ - odd ;}\\
\ss{0}{W}+\ss{4}{W}+\ldots \ss{4k-4}{W}, & \mbox{\rm $k\leq \frac{n}{2}$, $k\leq \frac{n}{4}+1$, $k$ - odd ;}\\
\ss{0}{W}+\ss{4}{W}+\ldots \ss{2n-4k+4}{W}, & \mbox{\rm $\frac{n}{4}+1\leq k \leq \frac{n}{2}$, $n$ - even, $k$ - odd ;}\\
\ss{0}{W}+\ss{4}{W}+\ldots \ss{2n-4k+2}{W}, & \mbox{\rm $\frac{n}{4}+1\leq k \leq \frac{n}{2}$, $n, k$ - odd ;}\\
\ss{0}{W}+\ss{4}{W}+\ldots \ss{4n-4k}{W}, & \mbox{\rm $\frac{3n}{4} \leq k$, $n, k$ are of same parity ;}\\
\ss{0}{W}+\ss{4}{W}+\ldots \ss{-2n+4k}{W}, & \mbox{\rm $\frac{n}{2}\leq k \leq \frac{3n}{4}$, $n, k$ - even ;}\\
\ss{0}{W}+\ss{4}{W}+\ldots \ss{-2n+4k-2}{W}, & \mbox{\rm $\frac{n}{2}\leq k \leq \frac{3n}{4}$, $n, k$ - odd ;}\\
\ss{0}{W}+\ss{4}{W}+\ldots \ss{4n-4k-4}{W}, & \parbox{.5\linewidth}{\rm $\frac{n}{2}\leq k$, $\frac{3n}{4}-1\leq k$,\\ $n,k$ are of different parity ;}\\
\ss{0}{W}+\ss{4}{W}+\ldots \ss{-2n+4k+4}{W}, & \mbox{\rm $\frac{n}{2}\leq k \leq \frac{3n}{4}-1$, $n$ - even, $k$ - odd ;}\\
\ss{0}{W}+\ss{4}{W}+\ldots \ss{-2n+4k+2}{W}, & \mbox{\rm $\frac{n}{2}\leq k \leq \frac{3n}{4}-1$, $n$ - odd, $k$ - even .}
\end{array}
\right.
$$
These and similar formulas are rather massive. So, we will not improve these formulas for any $m$. But we note that formulas in Theorem 6.2 can be improved for $mk > n$.
\end{rem}

\section{Some elementary functions of Clifford algebra elements}

Consider some elementary functions of Clifford algebra elements.

The following element
$$\exp U= e + U + \frac{(U)^2}{2!} + \frac{(U)^3}{3!}+ \ldots=\sum_{j=0}^{\infty} \frac{(U)^j}{j!}$$
is called {\it an exponent of Clifford algebra element} $U\in\cl(p,q)$.

In the same way we denote {\it sine, cosine, hyperbolic sine and hyperbolic cosine of Clifford algebra element}:
$$\sin \,U=\sum_{j=0}^{\infty} (-1)^j\frac{(U)^{2j+1}}{(2j+1)!},\quad
\cos \,U=\sum_{j=0}^{\infty} (-1)^j\frac{(U)^{2j}}{(2j)!},$$
$$\sinh \,U=\sum_{j=0}^{\infty} \frac{(U)^{2j+1}}{(2j+1)!},\quad
\cosh \,U=\sum_{j=0}^{\infty} \frac{(U)^{2j}}{(2j)!}.$$
These power series converge for any element $U$ of Clifford algebra. We know rank of any Clifford degree of Clifford algebra element. Then we can find rank of written out functions of Clifford algebra element. For example,
$$\exp\ss{0}{U}=e+ue+\frac{u^2}{2}e+\frac{u^3}{3!}e+\ldots=(\sum_{k=0}^{\infty}\frac{u^k}{k!})e\in \cl_{0}(p,q),$$
$$\exp\ss{1}{U}=\sum_{j=0}^{\infty}\frac{(\sum_{i=1}^{n}({u_i}^2\eta^{ii})^j)}{(2j)!}e+$$
$$+\sum_{j=0}^{\infty}\frac{(\sum_{i=1}^{n}({u_i}^2\eta^{ii})^j)}{(2j+1)!}\sum_{k=1}^n u_k e^k\in \cl_0(p,q)+\cl_1(p,q).$$
Since these formulas, it follows that
$$\sin\ss{0}{U},\ \cos\ss{0}{U},\ \sinh\ss{0}{U},\ \cosh\ss{0}{U} \in \cl_0(p,q),$$
$$\sin\ss{1}{U},\ \sinh\ss{1}{U} \in \cl_1(p,q),$$
$$\cos\ss{1}{U},\ \cosh\ss{1}{U} \in \cl_0(p,q).$$
Using Theorem 6.2, we get for $\overline k=\overline 0, \overline 1, \overline 2,\overline 3$:
$$\exp\ss{\overline k}{U}=\ss{\overline{0k}}{W},\quad
\sin\ss{\overline k}{U}=\ss{\overline k}{W},\quad
\sinh\ss{\overline k}{U}=\ss{\overline k}{W},\quad \cos\ss{\overline k}{U}=\ss{\overline 0}{W},\quad
\cosh\ss{\overline k}{U}=\ss{\overline 0}{W}.$$

In the same way let's denote {\it exterior exponent, exterior sine, exterior cosine, exterior hyperbolic sine, hyperbolic cosine of Clifford algebra element}:
$$ \ss{\wedge}{\exp}\, U= e + U + \frac{(\wedge U)^2}{2!} + \frac{(\wedge U)^3}{3!}+ \ldots=\sum_{j=0}^{\infty} \frac{(\wedge U)^j}{j!},$$
$$\ss{\wedge}{\sin}\, U=\sum_{j=0}^{\infty} (-1)^j\frac{(\wedge U)^{2j+1}}{(2j+1)!},\quad
\ss{\wedge}{\cos}\, U=\sum_{j=0}^{\infty} (-1)^j\frac{(\wedge U)^{2j}}{(2j)!},$$
$$\ss{\wedge}{\sinh}\, U=\sum_{j=0}^{\infty} \frac{(\wedge U)^{2j+1}}{(2j+1)!},\quad
\ss{\wedge}{\cosh}\, U=\sum_{j=0}^{\infty} \frac{(\wedge U)^{2j}}{(2j)!}.$$
Using Theorem 6.1, we get
$$\ss{\wedge}{\exp}\,\ss{0}{U}=e+ue+\frac{u^2}{2}e+\frac{u^3}{3!}e+\ldots=(\sum_{k=0}^{\infty}\frac{u^k}{k!})e\in \cl_{0}(p,q),$$
$$\ss{\wedge}{\sin}\,\ss{0}{U},\ \ss{\wedge}{\cos}\,\ss{0}{U},\ \ss{\wedge}{\sinh}\,\ss{0}{U},\ \ss{\wedge}{\cosh}\,\ss{0}{U} \in \cl_0(p,q),$$
$$
\ss{\wedge}{\exp}\,\ss{k}{U}=\left\lbrace
\begin{array}{ll}
e+\ss{k}{U} \in \cl_0(p,q)+\cl_k(p,q), & \mbox{\rm $k$ - odd ;}\\
\ss{0}{W}+\ss{k}{W}+\ss{2k}{W}+\ss{3k}{W}+\ldots, & \mbox{\rm $k$ - even,}
\end{array}
\right.
$$
$$
\ss{\wedge}{\sin}\,\ss{k}{U},\ \ss{\wedge}{\sinh}\,\ss{k}{U}=\left\lbrace
\begin{array}{ll}
\ss{k}{U} \in \cl_k(p,q), & \mbox{\rm $k$ - odd ;}\\
\ss{k}{W}+\ss{3k}{W}+\ss{5k}{W}+\ss{7k}{W}+\ldots, & \mbox{\rm $k$ - even,}
\end{array}
\right.
$$
$$
\ss{\wedge}{\cos}\,\ss{k}{U},\ \ss{\wedge}{\cosh}\,\ss{k}{U}=\left\lbrace
\begin{array}{ll}
e \in \cl_0(p,q), & \mbox{\rm $k$ - odd ;}\\
\ss{0}{W}+\ss{2k}{W}+\ss{4k}{W}+\ss{6k}{W}+\ss{8k}{W}+\ldots, & \mbox{\rm $k$ - even.}
\end{array}
\right.
$$

\section{Conclusion}

The method of quaternion typification in his primary form \cite{quattyp} allowed us to find quaternion type of commutators and
anticommutators of Clifford algebra elements with given quaternion types. In this paper we have improved the method. Now we know quaternion type of all expressions $$[U_1, U_2, \ldots, U_k]=U_1 U_2 \ldots U_k-U_k \ldots U_2 U_1,$$ $$\{U_1, U_2, \ldots, U_k\}=U_1 U_2 \ldots U_k+U_k \ldots U_2 U_1,$$ and $$U_1 U_2 \ldots U_k$$  where $k>1$ - integer number.

Also we know quaternion type of different other expressions as Clifford and exterior degrees, elementary functions of Clifford algebra elements.

In many cases, a classification of Clifford algebra elements according to their quaternion types makes it possible to obtain instructive results having no rank analogues.

\end{document}